\documentclass[onecolumn,aps,showpacs,floatfix,superscriptaddress]{revtex4}
\usepackage{amsmath,amsfonts,amssymb,eucal,graphicx,graphics,bm}
\newcommand{\rmd}{\mathrm{d}} %
\newcommand{\eref}[1]{Eq.~(\ref{#1})}%
\newcommand{\fref}[1]{Fig.~\ref{#1}} %

\begin{document}
\title{Crowding at the Front of the Marathon Packs} \author{Sanjib~Sabhapandit}
\email{sanjib.sabhapandit@u-psud.fr} \affiliation{Laboratoire de Physique
  Th\'eorique et Mod\`eles Statistiques (UMR 8626 du CNRS), Universit\'e
  Paris-Sud, B\^atiment 100, 91405 Orsay Cedex, France}
\author{Satya~N.~Majumdar} \email{majumdar@lptms.u-psud.fr}
\affiliation{Laboratoire de Physique Th\'eorique et Mod\`eles Statistiques
  (UMR 8626 du CNRS), Universit\'e Paris-Sud, B\^atiment 100, 91405 Orsay
  Cedex, France} \author{S.~Redner} \email{redner@bu.edu}
\affiliation{Center for Polymer Studies and Department of Physics, Boston
  University, Boston, Massachusetts 02215 USA}

\begin{abstract}
  We study the crowding of near-extreme events in the time gaps between
  successive finishers in major international marathons.  Naively, one
  might expect these gaps to become progressively larger for better-placing
  finishers.  While such an increase does indeed occur from the middle of
  the finishing pack down to approximately $20^{\rm th}$ place, the gaps
  saturate for the first 10--20 finishers.  We give a probabilistic account
  of this feature.  However, the data suggests that the gaps have a weak
  maximum around the $10^{\rm th}$ place, a feature that seems to have a
  sociological origin.

\end{abstract}
\maketitle

It is fun to learn about sports statistics and discuss their implications
among fellow sports fans.  The existence of comprehensive web-based resources
for sports statistics, whose easy availability was unimaginable just a few
years ago, has perhaps helped promote such activities.  In this note, we
investigate one such statistic, namely, the finishing times of individual
runners in major marathons \cite{data}.  Our main interest is in the
dependence of the {\em time gaps between successive finishers} on finishing
place.  More precisely, let $t_k$ be the time of the $k^{\rm th}$ finisher.
Then we wish to understand how the time gaps $g_k\equiv t_{k+1} -t_k$ depend
on finishing place $k$.  Because front runners are rare and potential race
leaders are rarer still, the natural expectation is that the gaps between
successive finishers should increase monotonically in moving from the middle
of the pack towards the increasingly-rare front runners.  However, the data
show that the time gaps saturate to a constant value for sufficiently small
$k$.  We suggest that sociological factors may contribute to this anomaly in
the gaps.

\begin{figure}[ht]
\centerline{  \includegraphics*[width=0.45\textwidth]{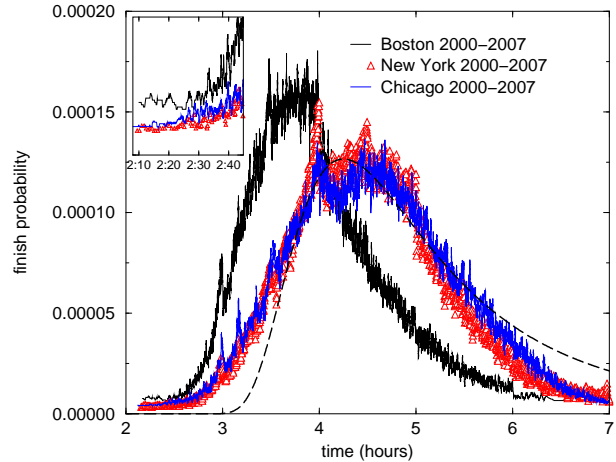}}
\caption{Distribution of all finishing times (smoothed over a 20-point range
  for visual clarity) for the Boston, Chicago, and New York marathons,
  2000--2007.  Notice the peaks at 3 hours in all the data, the prominent
  peaks at 4 hours for the Chicago and New York marathons, and the secondary
  peaks at 3:10, 3:20, and 3:30 for the Chicago marathon.  The dashed curve
  shows the distribution of Eq.~\eqref{Pte}, with parameter values as given
  in the text.  The inset shows the data in the range of 2:08--2:45.  }
  \label{finish}
\end{figure}

The results presented here are based on data for finishing times in major
international marathons that attract world-class entrants.  These include
Boston, Chicago, and New York from 2000--2007 (entire fields), as well as
Berlin 1992 and 1999--2007, Fukuoka 2006--2007, London 2001--2007, and Paris
2004, 2006--2007 (first 100 places for all non-US races).  Data for other
years in these non-US marathons is not readily available or corrupted, and
some of the data used in this work required corrections of a few obviously
erroneous results.  In these marathons, the winning time is in the range
2:05--2:10.  For example, in Boston, Chicago, and New York, the course
records are 2:07:14, 2:05:42, and 2:07:43, respectively, while the current
world record, set by Haile Gebrselassie in the 2007 Berlin Marathon, is
2:04:26.  After the race winner, there is a trickle of fast finishers that
gradually turns into a steady flow as the finish time approaches 3 hours.
The main pack arrives in the range of 3--6 hours, with a decreasing stream of
progressively slower stragglers.  Thus one naturally anticipates the
distribution of finish times shown in Fig.~\ref{finish}.

Upon examining these distributions critically, a number of curiosities can be
seen.  First, in spite of the data smoothing, there are visible peaks at just
under 3 hours and 4 hours for all three marathons.  For the Chicago marathon
in particular, where the course is flat and well-suited for pacing, one can
even discern secondary peaks near 3:10, 3:20, and 3:30 (Fig.~\ref{finish}).
The existence of such peaks suggests that the distribution of finish times in
this range does not reflect a performance limit, but rather, the surmounting
of a psychological barrier.  Parenthetically, the apparent difference in the
distributions for the Boston marathon (where challenging qualifying times
exist), with the Chicago and New York marathons can be made to nearly
disappear by plotting them in scaled units---namely, by making the abscissa
the finish time divided by the average finish time for each set of 8 races.

\begin{figure}[ht]
\centerline{\includegraphics*[width=0.5\textwidth]{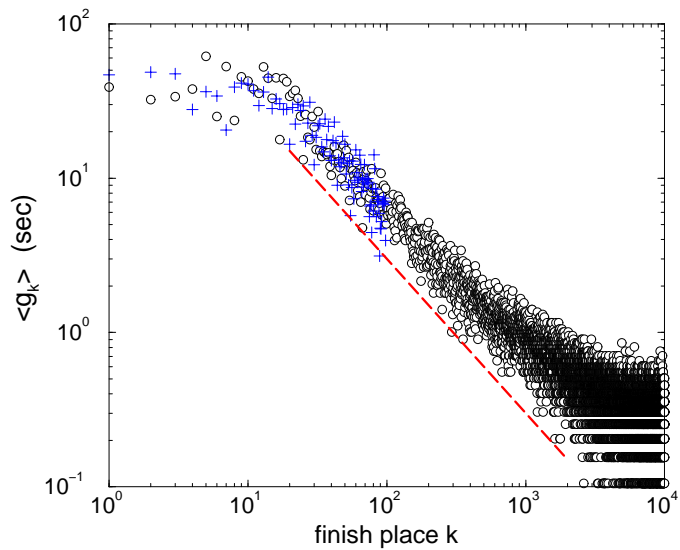}}
\caption{Distribution of the average time gap $g_k$ between the $k^{\rm th}$
  and $(k+1)^{\rm st}$ finisher for: the US ($\circ$) and the European ($+$)
  marathons cited in the text.  For the US marathons the first 10,000 gaps
  are shown, while the first 100 gaps are shown for the European marathons.
  The dashed line has a slope of $-1$, as given by Eq.~\eqref{dTk}.}
  \label{dt-av}
\end{figure}

More interesting behavior, and the main point of this work, is the $k$
dependence of the time gaps $g_k$ between successive finishers.  We are
particularly interested in these gaps for finishers near the front of the
pack.  Thus we restrict ourselves to the first 10,000 finishers in the US
marathons.  This threshold corresponds to finishing times of about 4 hours
for Chicago and New York, and around 3:45 for Boston.  By comparing with
Fig.~\ref{finish}, these time thresholds are prior to the peak of the
finishing time distribution for Chicago and New York, and near the peak for
Boston.  For comparison, the average number of finishers over the last eight
US marathons that we studied is 30,668 for Chicago, 33,669 for New York, and
16,645 for Boston.  For $k>10,000$, $\langle g_k\rangle$ begins increasing,
corresponding to the lagging tail of the finishing time distribution.  For
the European data, we quote $g(k)$ only to $k=100$.

Among the fastest finishers, the finish time distribution decays very slowly
and is nearly constant for times less than 2:30 (inset to Fig.~\ref{finish}).
For the marathons that we studied, the {\em average} time gap between
consecutive finishers among the first 10 places is in the range of 20--60
seconds, and do not have any clear systematic $k$ dependence
(Fig.~\ref{dt-av}).  Members of this group of elite runners are all possible
candidates to win the race on any given day.  In contrast, beyond the
$20^{\rm th}$ place, the average gap systematically decreases with $k$, a
decrease that clearly reflects the increase in the density of runners as the
leading edge of the pack arrives at the finish.

We can make these observations quantitative by assuming that the finishing
times of individual runners are independent and identically distributed (iid)
random variables, and then using extreme-value statistics to determine the
time gaps $g_k$ between successive finishers \cite{extremal,gumbel}.  As a
preliminary, consider the time $t_k$ of the $k^{\rm th}$ finisher.  The {\em
  typical\/} value for this time can be determined from the extremal
condition (which assumes self averaging)
\begin{equation}
\label{ext}
  \int_0^{t_k}  P(t)\, dt \approx \frac{k}{N}~,
\end{equation}
that states that there are $k$ individuals whose finishing times are less
than $t_k$.  The resulting estimate for the typical $k^{\rm th}$ finishing
time $t_k$ should be accurate for $k\gg 1$, where fluctuations in $t_k$ are
negligible.  More generally, we can compute the full probability distribution
of $t_k$, as outlined in appendix~\ref{pdf t_k} and thereby find the mean
value of $t_k$ to be
\begin{equation}
\label{I_y}
  \langle t_k \rangle = \int_0^\infty
  I\left(P_>(x);N\!-\!k\!+\!1,k\right)\,\rmd x\,, 
\end{equation}
where 
\begin{math}
I(y;a,b)= \bigl[\int_0^{y} x^{a-1}\,(1-x)^{b-1}\, \rmd x\bigr]\big/\bigl[\int_0^1
x^{a-1}\,(1-x)^{b-1}\, \rmd x\bigr]
\end{math}
is the regularized (in the sense that $I(1;a,b)=1$) incomplete Beta function
and $P_>(x)\equiv \int_x^\infty P(x')\, \rmd x'$ is the exceedance
probability.

The main message from either the exact result in \eref{I_y} or the extremal
condition in \eref{ext}, is that the time gap $ g_k = t_{k+1} - t_{k}$, has
the following generic behaviors (see appendix~\ref{appendix} for three simple
examples):

\begin{itemize}
\itemsep -3pt

\item If $P(t)$ is constant, then $\langle g_k\rangle$ is independent of $k$.

\item If $P(t)$ increases monotonically as $t$ increases, then $\langle
  g_k\rangle$ decreases monotonically as $k$ increases.

\item If $P(t)$ decreases monotonically as $t$ increases, then $\langle
  g_k\rangle$ increases monotonically as $k$ increases.
\end{itemize}

Let us now apply the above results to marathon finishing times.  As a trivial
and artificial initial example, suppose that the marathon runners' speeds $s$
are distributed exponentially, $P(s)= s_*^{-1}\, e^{-s/s_*}$, with $s_*$ a
characteristic running speed.  Then the distribution of finishing times
$t=L/s$ would be
\begin{equation}
\label{Pt}
P_0(t)= \frac{T}{t^2}\,\,e^{-T/t}~,
\end{equation}
where $T=L/s_*$ is a typical finishing time for the field, and $L$ is the
course length.  Applying the extremal criterion \eqref{ext} to this
distribution gives the typical $k^{\rm th}$ finishing time $t_k=
T/[\ln(N/k)]$.  While $t_k$ increases with $k$, as it must, this result has
the unrealistic feature that the winning time approaches zero as the field
becomes arbitrarily large.

More plausibly, the finishing time distribution should incorporate a non-zero
fastest time $t_{\rm min}$.  A slightly more refined example that obeys this
constraint is
\begin{equation}
\label{Pte}
P(t)= \frac{mT^m}{\tau^{m+1}}\,\,e^{-(T/\tau)^m}~,
\end{equation}
where $\tau =(t-t_{\rm min})$.  The main new features of this distribution
compared to Eq.~\eqref{Pt} are the cutoff at $t_{\rm min}$ and the arbitrary
exponent value $m$; the power-law prefactor is subdominant and it merely
serves to simplify the calculations below.  In fact, with the values $t_{\rm
  min}=1.75$ hours, $T=2.75$ hours and $m=3$, Eq.~\eqref{Pte} roughly follows
the data in Fig.~\ref{finish} (dashed curve).  While one should not take the
distribution \eqref{Pte} and the parameter values too seriously, we will see
that its precise form does not affect the behavior of the time gaps between
successive finishers.

Applying the extremal condition \eqref{ext} to the distribution \eqref{Pte},
and using the variable change $x=(T/\tau)^m$ to simplify the resulting
integral, the typical value of the time gap is
\begin{eqnarray}
\label{dTk}
g_k = T\left[ \left(\ln\frac{N}{k+1}\right)^{-1/m}- 
  \left(\ln\frac{N}{k}\right)^{-1/m}\right]
~\approx~ \frac{T}{m(\ln N)^{1+1/m}}\,\,\, \frac{1}{k}\, \quad1\ll k\ll N.
\end{eqnarray}
This $1/k$ dependence holds for any distribution with an exponentially fast
cutoff near the lower limit.  The behavior $g_k\propto 1/k$ accords well with
the data beyond approximately $20^{\rm th}$ place.  However, contrary to the
prediction of Eq.~\eqref{dTk}, the data clearly show that there is an
``excess'' of elite runners (Fig.~\ref{dt-av}), as the time gaps between
successive finishers are roughly constant for the first 20 places.  Moreover,
for the US races, the gaps between the first few consecutive finishers
actually decrease with $k$.  As seen in Fig.~\ref{dt-av} for US races, the
largest gap occurs between $5^{\rm th}$ and $6^{\rm th}$ place.

The reason that \eref{dTk} does not capture the small-$k$ behavior seen in
\fref{dt-av} is that the parent distribution in \eref{Pte} quickly goes to
zero close to the fastest finishing time $t_\text{min}$, whereas the actual
distribution becomes nearly flat in this regime (\fref{finish}).  If we were
to consider a flat distribution $P(t)$, as suggested by the data shown in
\fref{finish}, then a constant gap would be reproduced.  The generic behavior
of the dependence of the gap $g_k$ on $k$ is discussed in
appendix~\ref{appendix}.  Along these lines, a recent theory \cite{density}
predicts a crowding of runners near the front of marathon packs when the
finishing time distribution is bounded from below.  One additional feature of
the gaps is that they begin to increase with $k\agt 1000$ (\fref{dt-av}).
This behavior also arises from \eref{dTk} for large $k$.  This regime
corresponds to finishing times of more than $4$ hours and is not relevant for
our main conclusions.

Is there an explanation for having an excess of world-class runners?  Many
elite runners enjoy considerable incentives to maintain their competitive
edge, including appearance money, access to the best support institutions
(medical and athletic), etc.  Thus if one achieves a time that qualifies as
an elite performance, one is then in a position to take advantage of the
various inducements offered to leading runners to maintain such a status.
However, runners at the next tier of achievement face a daunting challenge.
To run a marathon in the range, say, of 2:15--2:30 (for men) is still an
impressive achievement that requires significant talent, dedication, and time
commitment.  However, such a finish time is too slow to be competitive at
major marathons.  Thus runners who finish in this range typically have little
or no external support for their athletic activities and have to balance this
all-consuming endeavor with the need to survive economically.  Consequently,
one may even anticipate a deficit of male runners who can complete a marathon
in the range of 2:15--2:30.  Such a feature does actually occur in the Boston
marathon.

It would be valuable to study whether a similar excess of elite exists in
different athletic events or other forms of human competition.  It is also
worth mentioning that perhaps a similar elite excess occurs in human
mortality, where there is a well-known mortality plateau among the
longest-lived individuals \cite{plateau,penna,azbel}.  Here again, there
seems to be a self-selected sub-population of advantaged individuals who gain
advantage both innately and perhaps because of external reinforcement.

{\bf Acknowledgments.}  One of us (SR) thanks Guoan Hu for his invaluable
data collection assistance, Paul Krapivsky for a helpful discussion, and
financial support from the NSF.  SS and SNM acknowledge the support of the
Indo-French Centre for the Promotion of Advanced Research under Project
3404-2.

\appendix

\section{\label{pdf t_k} Probability distribution of the $k^\text{th}$
  finishing time}

For a set of $N$ iid random times that are drawn from the same
distribution $P(t)$, let $\{t_1, t_2,\dotsc,t_N\}$ denote their ordered set,
with $t_1 < t_2< \dotsb <t_N$.  Thus $t_1$ denotes the winning time, $t_2$
denotes the $2^\text{nd}$ place time, and $t_k$ denotes the $k^\text{th}$
fastest finishing time.

The probability distribution of the $k^\text{th}$ fastest finishing time
$t_k$ is given by
\begin{eqnarray}
\label{f(t_k)}
f(t_k)&=&\frac{N!}{ (N-k)!(k-1)!} 
\left[\int_{t_k}^\infty P(x)\, \rmd x \right]^{N-k}
\left[\int_{0}^{t_k} P(x)\, \rmd x \right]^{k-1} 
P(t_k),  \nonumber  \\
&=&\frac{1}{B(N-k+1,k)}
\Bigl[P_>(t_k)\Bigr]^{N-k}\, \Bigl[1-P_>(t_k)\Bigr]^{k-1}\,
 \left[-\frac{\rmd P_>(t_k)}{\rmd t_k}\right]\,.
\end{eqnarray}
Equation~\eqref{f(t_k)} merely specifies that $(N-k)$ variables are greater
than $t_k$, $(k-1)$ variables are smaller than $t_k$, and one variable equals
$t_k$.  The combinatorial prefactor gives the number of such arrangements of
these variables.  In the second line, $B(a,b)=\Gamma(a)\Gamma(b)/\Gamma(a+b)$
is the Beta function, and we have defined the exceedance probability
\begin{equation*}
P_>(x)\equiv \int_x^\infty P(x')\, \rmd x',
\end{equation*}
namely, the probability that a variable chosen from the initial distribution
$P$ exceeds $x$.  This exceedance probability satisfies the obvious
conditions $P_>(0)=1$ and $P_>(\infty)=0$.  One can easily check from
\eref{f(t_k)} that $f(t_k)$ is normalized, {\it i.e.}, $\int_0^\infty
f(t_k)\, \rmd t_k =1$, as it must.

The average value of the $k^\text{th}$ fastest finishing time is then
\begin{eqnarray}
\label{t_k 2}
\langle t_k \rangle&=& \frac{1}{B(N\!-\!k\!+\!1,k)}\int_0^\infty x\,\,
\Bigl[P_>(x)\Bigr]^{N-k} \Bigl[1-P_>(x)\Bigr]^{k-1}
 \left[-\frac{\rmd P_>(x)}{\rmd x}\right]\,\rmd x\nonumber \\ \nonumber\\
&\equiv &- \int_0^\infty x\,\, \frac{\rmd I}{\rmd x}\,\, dx.
\end{eqnarray}
In the second line we have introduced $I=I(y;a,b)$, the regularized
incomplete Beta function, $I(y;a,b)\equiv B(y;a,b)/B(a,b)$, in which $B(y;a,b)$
the incomplete Beta function
\begin{equation*}
B(y;a,b)=\int_0^y x^{a-1} (1-x)^{b-1}\, \rmd x, \qquad \qquad y\in[0,1]\,,
\end{equation*}
$B(a,b)=B(1;a,b)$ the standard Beta function, and $y\equiv P_>(x)$.
Integrating \eref{t_k 2} by parts, and using the fact that the integrated
term vanishes at both endpoints, gives the mean $k^\text{th}$ finishing time
expressed by \eref{I_y}.

\section{\label{appendix} $\langle g_k\rangle$ for three simple cases}

In this appendix we calculate $\langle g_k\rangle$ explicitly for three
simple cases of $P(t)$.
\medskip

\emph{Case 1.} For the uniform distribution $P(t)=1$ in $t\in[0,1]$ and
$P(t)=0$ outside. Hence,
$P_>(t)=1-t$.  Then
\begin{align}
\langle t_k\rangle &=\frac{1}{B(N-k+1,k)}\int_0^1 x\cdot x^{k-1} (1-x)^{N-k}\,
  dx = \frac{B(N-k+1,k+1)}{B(N-k+1,k)} \nonumber \\
& = \frac{k}{N+1}~.
\end{align}
Thus we obtain $\langle g_k\rangle=1/(N+1)$ for all $k$, while using the
extremal condition \eref{ext} one finds the typical gap $g_k\approx 1/N$. As
expected, $\langle g_k\rangle$ is independent of $k$ for a uniform
distribution.  \medskip

\emph{Case 2.} Consider the monotonically increasing distribution $p(t)=2 t$
in $t\in[0,1]$. Then $P_>(t)=1-t^2$. Hence,
\begin{align}
\langle t_k\rangle &=\frac{1}{B(N-k+1,k)}\int_0^1 x\cdot x^{2(k-1)} (1-x^2)^{N-k}\,
  dx \nonumber = \frac{1}{B(N-k+1,k)}\int_0^1 z^{k-1/2} (1-z)^{N-k}\,dz
  \nonumber \\ \nonumber\\
&  = \frac{B(N-k+1,k+1/2)}{B(N-k+1,k)}= \frac{\Gamma(N+1)\Gamma(k+1/2)}{\Gamma(k)\Gamma(N+3/2)}~.
\end{align}
From this exact calculation we find
\begin{align}
\langle g_k\rangle=\frac{\Gamma(N+1)}{\Gamma(N+3/2)}\cdot
\frac{\Gamma(k+1/2)}{\Gamma(k)} \cdot \frac{1}{2k}, \quad\text{for
  all}~N \quad\text{and}\quad \approx \frac{1}{2\sqrt N}\cdot \frac{1}{\sqrt k},
\quad\text{for}~ N>k\gg 1.
\end{align}
Similarly using \eref{ext} the typical value of the $k^{\rm th}$ finishing
time is $t_k \approx \sqrt{k/N}$, and hence
\begin{align}
g_k&\approx \frac{1}{\sqrt N} \left[\sqrt{k+1}-\sqrt{k}\right] \nonumber \\
&\approx \frac{1}{2\sqrt N}\cdot \frac{1}{\sqrt k}, \quad\text{for}~ N>k\gg 1.
\end{align}
These results show that $\langle g_k\rangle$ monotonically decreases as $k$
increases.  That is, the gap between successive variables gets smaller when
their density increases, as one would expect.
\medskip

\emph{Case 3.}  Consider the monotonically decreasing distribution
$P(t)=\exp(-t)$ where $t\in[0,\infty)$.  In this case,
\begin{align}
\langle t_k\rangle &=\frac{1}{B(N-k+1,k)}\int_0^\infty x\cdot e^{-(N-k)x} (1-e^{-x})^{k-1}\,
  dx \nonumber = - \frac{1}{B(N-k+1,k)}\int_0^1 {\ln z}^{N-k} (1-z)^{k-1}\,dz\,.
\end{align}
The latter integral can be found in Gradshteyn and Ryzhik \cite{GR} and the final
result is
\begin{equation}
\langle t_k\rangle = \psi(N+1)-\psi(N-k+1)\,,
\end{equation}
where $\psi(x) =\frac{d \ln\Gamma(x)}{dx}$ is the digamma function.  Finally
using the series representation
\begin{equation}
\psi(n) = -\gamma+\sum_{m=1}^{n-1}\frac{1}{m},
\end{equation}
where $\gamma =0.577215\ldots$ is Euler's constant, we obtain
\begin{align}
\langle g_k\rangle=\frac{1}{N-k}, \quad 1\le k\le N-1.
\end{align}
On the other hand using the extremal condition \eref{ext} one finds the typical value
\begin{align}
g_k \approx -\log \left(1-\frac{1}{N-k} \right) \approx\frac{1}{N-k},
\quad\text{for}~N-k \gg 1.
\end{align}
Thus $\langle g_k\rangle$ monotonically increases as $k$ increases.  Note
that in this case the extremal condition \eref{ext} does not describe
well the behavior when $k$ is close to $N$.

\end{document}